\documentclass[iop,apj]{emulateapj}

\usepackage{bm} 
\usepackage{amsmath} 
\usepackage{hyperref} 
\usepackage{natbib} 
\usepackage{textcomp} 
\usepackage{graphicx} 

\newcommand{\sdata}{$N_{\odot,r}$} 
\newcommand{\nggn}{$(n,\gamma)\rightleftarrows(\gamma,n)$}  
\newcommand{\reptaungafewb}{$\tau^{REP}_{n\gamma}\approx\text{ a few } \tau^{REP}_{\beta}$}     

\begin{document}

\title{The Rare Earth Peak : An Overlooked r-Process Diagnostic}

\author{Matthew R. Mumpower}
\email{mrmumpow@ncsu.edu}
\affil{Department of Physics, North Carolina State University, Raleigh, North Carolina 27695-8202,USA}

\author{G. C. McLaughlin}
\email{gail\_mclaughlin@ncsu.edu}
\affil{Department of Physics, North Carolina State University, Raleigh, North Carolina 27695-8202,USA}

\and

\author{Rebecca Surman}
\email{surmanr@union.edu}
\affil{Department of Physics and Astronomy, Union College, Schenectady, New York 12308,USA}

\begin{abstract}
The astrophysical site or sites responsible for the $r$-process of nucleosynthesis still remains an enigma. Since the rare earth region is formed in the latter stages of the $r$-process it provides a unique probe of the astrophysical conditions during which the $r$-process takes place. We use features of a successful rare earth region in the context of a high entropy $r$-process ($S\gtrsim100k_B$) and discuss the types of astrophysical conditions that produce abundance patterns that best match meteoritic and observational data. Despite uncertainties in nuclear physics input, this method effectively constrains astrophysical conditions.
\end{abstract}



\section{Introduction}\label{intro}
Among many open problems in modern nuclear astrophysics, a complete understanding of the rapid neutron capture or ``$r$-process'' remains one of the greatest challenges. The $r$-process is believed to be responsible for approximately half the heavy elements above the iron group. There are many promising candidate sites for this synthesis process, however the exact astrophysical site has not yet been identified; for reviews see \citep{Arnould2007,Qian2007}. Some examples of candidate sites include high-entropy supernovae ejecta \citep{Meyer1992,Qian1996,Sumiyoshi2000,TBM2001}, compact object mergers \citep{Lattimer1977,Meyer1989,Freiburghaus1999,Goriely2005,Surman2008,Metzger2010}, neutrino induced nucleosynthesis in He shells \citep{Woosley1990,Banerjee2011}, or possibly $\gamma$-ray burst scenarios \citep{Surman2004,McLaughlin2005,Surman2006}. 

The study of the $r$-process is further complicated by the short lived nuclides far from stability. Recent experiments have led to measurements on some nuclides participating in the $r$-process e.g. \citet{Hosmer2005,Matos2008,Rahaman2008,Baruah2008,Jones2009,Hosmer2010,Nishimura2011,Quinn2011}. However current experimental data on neutron-rich isotopes is sparse; see e.g. figure 12 in \citet{Mumpower2011}. Therefore, theoretical extrapolations must be used as no experimental data are available \citep{Grawe2007}. Much theoretical work is on going to understand the properties of nuclei far from stability and their impact on the $r$-process, see e.g. \citet{Moller1995,ETFSIQ1996,NONSMOKER1998,ADANDT2000,Moller2003,Beun2008,Surman2009,HFB17,Mumpower2011,Arcones2011b,Suzuki2012}.

\citet{SuessUrey1956} demonstrated the existence of the $A=130$ and $A=195$ peaks in an improved distribution of solar abundances. The prominent abundance features seen in this curve were quickly attributed closed neutron shells occurring at magic neutron numbers \citep{BBFH}. The rare earth peak ($A\sim160$), lying away from closed neutron shells, must occur by a different mechanism.

Three mechanisms have been proposed for the formation of the rare earth peak. (1) It has been hypothesized that this peak may form by fission cycling if the distribution of fission fragments in $(A,Z)$ is favorable \citep{Cameron1957,Schramm1971,MartiZeh1985,Goriely2011}. (2) \citet{Surman1997} discovered that the rare earth peak can form during freeze-out as material decays back to stability. Crucial to this argument was that this mechanism occur in a `hot freeze-out', with temperature high enough to support significant photo-dissociation flow. This allowed material to `funnel' into the peak in an attempt to sustain \nggn \ equilibrium. \citet{Otsuki2003} found similar abundance patterns provided the temperature was constant during freeze-out. (3) A third mechanism has also been found for lower temperature environments where a `cold freeze-out' may occur \citep{Mumpower2011}. In this scenario the temperature dependent photo-dissociations flow is not present, however the rare earth peak may still form by a trapping mechanism involving slow neutron capture rates in the region. The formation process, in addition to being sensitive to freeze-out conditions was also shown to be sensitive to nuclear physics inputs \citep{Mumpower2011,Arcones2011a}.

One of the most intriguing results to date regarding $r$-process yields is that the stable elements observed in galactic halo stars (much older than the sun) with $56\lesssim Z\lesssim80$, including the rare earth elements ($Z=57$ to $Z=71$), are consistent with the scaled solar $r$-process elemental abundances e.g. see \citet{CS22892}. This is fascinating because it suggests these elements were formed by an $r$-process mechanism which operates over a wide range of metallicity, e.g. see \citet{Cowan1997}.

An understanding of the production of the rare earth elements can be used to probe the freeze-out conditions of the $r$-process mechanism(s) which operates over a wide range of metallicity. This approach complements previous calculations which have focused on the conditions at the onset of the $r$-process i.e. the neutron-to-seed ratio \citep{HWQ1997,MeyerBrown1997,Freiburghaus1999}.

In this paper, we introduce three new rare earth peak constraints on astrophysical conditions in the context of the site of the $r$-process. Using the formation of the rare earth peak as our key constraint, we isolate freeze-out conditions which are favorable for the production of these elements. We compare final abundances from simulations to both the elemental abundances of halo stars and the isotopic abundances of $r$-process residuals. It is essential to note that (1) the $r$-process abundance pattern we observe may come about from a superposition of abundance patterns made from different conditions in the same astrophysical environment and (2) there exists significant variation between the final abundances using different nuclear models. We take this into consideration by averaging over and computing the variance of final abundances given different nuclear models and conditions. This allows us to compare final abundances, with error bars, to meteoritic and observational data.

\section{r-Process Model}\label{model}
Nuclides decay back to stability during freeze-out, the last stage of the $r$-process. To monitor this progress, it is useful to consider abundance weighted timescales for the primary reaction channels: neutron capture, photo-dissociation and $\beta$-decay.

\begin{subequations}\label{eqn:taus}
\begin{equation}
\label{eqn:tau-ng} \tau_{n\gamma} \equiv \frac{\sum_{Z\geqslant8,A}Y(Z,A)}{\sum_{Z\geqslant8,A}N_{n}\langle \sigma v\rangle_{Z,A} Y(Z,A)} 
\end{equation}
\begin{equation}
\label{eqn:tau-gn} \tau_{\gamma n} \equiv \frac{\sum_{Z\geqslant8,A}Y(Z,A)}{\sum_{Z\geqslant8,A}\lambda_{\gamma n}(Z,A)Y(Z,A)} \\
\end{equation}
\begin{equation}
\label{eqn:tau-beta} \tau_{\beta} \equiv \frac{\sum_{Z\geqslant8,A}Y(Z,A)}{\sum_{Z\geqslant8,A}\lambda_{\beta}(Z,A)Y(Z,A)}
\end{equation}
\end{subequations}

In equation \ref{eqn:taus}, the neutron number density is denoted by $N_{n}$. The Maxwellian averaged neutron capture cross section for nuclide $(Z,A)$ is $\langle \sigma v\rangle_{Z,A}$ and $\lambda_{\gamma n}(Z,A)$ the photo-dissociation rate. The full $\beta$-decay rate (including $\beta$-delayed neutron emission channels) for nuclide $(Z,A)$ is denoted by $\lambda_{\beta}(Z,A)$ and $Y(Z,A)$ is the abundance of nuclei $(Z,A)$. A reduced sum over the rare earth peak region ($A=150$ to $A=180$), denoted with a superscript ``REP,'' may be taken when applicable.

The neutron-to-seed ratio or $R$ is another useful metric, we define as:

\begin{equation}\label{eqn:r}
R \equiv \frac{Y_{n}}{\sum_{Z\geqslant8,A}Y(Z,A)}
\end{equation}

where $Y_{n}$ is the abundance of free neutrons. We consider \textquoteleft low\textquoteright \ neutron-to-seed ratio to be the time after neutron exhaustion, or $R=1$.

The final abundances are highly dependent on the rate of decline in the temperature and density during the last stage of the $r$-process \citep{Arcones2011a,Mumpower2011}. Among many possibilities, two thermodynamic evolutions are noteworthy and could occur in any astrophysical environment. A \textquoteleft hot\textquoteright\ freeze-out which proceeds under high temperatures ($T_{9}\gtrsim0.6$) when material decays back to stability and a \textquoteleft cold\textquoteright\ freeze-out which proceeds under low temperatures ($T_{9}\lesssim0.6$) when material decays back to stability, e.g. \citet{Wanajo2007}.

We contrast the differences in the evolution of the rare earth region between the two thermodynamic trajectories in Figure \ref{fig:timeline}. The trajectories used in this figure are from \citet{Mumpower2011} and the values in the timeline represent typical simulations which produce a satisfactory rare earth region compared to the solar isotopic abundances. We note that all of the simulations studied here shown qualitatively similar behavior. For further discussions of late time $r$-process dynamics see \citep{Surman1997,Surman2001,Arcones2011a,Mumpower2010,Mumpower2011}.

The rare earth region typically breaks from \nggn \ equilibrium at $T_9\sim1$, represented by point A in the figure. In the hot evolution, this is due to neutron exhaustion ($R=1$) while in the cold evolution this is due to the rapidly dropping density. Once the neutron-to-seed-ratio is unity, point B, the rare earth peak may begin to form \citep{Surman1997,Mumpower2011}. This process usually lasts until $\beta$-decays take over neutron captures in the region, point C. However, small changes in the pattern can continue after point C, until \reptaungafewb (not shown in the figure). The abundance averaged neutron separation energy is denoted by $\langle S_n\rangle$. Following this quantity along the timeline we clearly see the rare earth peak forms as the nuclides decay back to stability (increasing values of $\langle S_n\rangle$) and that cold freeze-out calculations typically extend farther from stability, closer to the neutron dripline than hot freeze-out calculations.

In order to probe the conditions which are ideal for the $r$-process as matter decays back to stability, we study a range of possible outflow parameterizations. Our model consists of a monotonically decreasing temperature with density parameterized as:

\begin{equation}\label{eqn:rho-full}
\rho(t)=\rho_{1}\text{exp}(-t/\tau)+\rho_{2}\left(\frac{\Delta}{\Delta+t}\right)^{n}
\end{equation}

In equation \ref{eqn:rho-full}, t denotes simulation time, $\rho_{1}+\rho_{2}$ is the initial density at time $t=0$, $3\tau=\tau_{dyn}$ with $\tau_{dyn}$ the dynamical timescale and $\Delta$ is a constant real number. For our calculations we take $\tau\sim27 \text{ms}$. The combination of $\tau$ and entropy per baryon, $S$, determines the distribution of seeds and the value of $R$ at the onset of the neutron capture phase. We hold $\tau$ constent and vary $S$ in our calculations. The parameter $n$ sets the thermodynamic behavior of the evolution at a low neutron-to-seed ratio. A hot $r$-process evolution has typical values of $n\sim2$ which is characteristic of wind models \citep{Meyer2002,Panov2009}. Values of $n\gtrsim5$ are typical of cold $r$-process evolutions and correspond to a faster decline \citep{Wanajo2007}.

Our $r$-process calculations are conducted with a nuclear reaction network which includes approximately 4000 nuclides relevant to the $r$-process. This code is fully dynamic in the sense that it does not assume any equilibrium conditions. The initial abundances for these nuclides are taken from an intermediate reaction network \citep{Hix1999} with PARDISO solver \citep{PARDISO}. During the decay back to stability the primary reaction channels are neutron capture, photo-dissociation, beta-decay, beta-delayed neutron emission, spontaneous fission and beta-delayed fission.

Asymmetric fission is included during these freeze-out calculations \citep{Seeger1965} and operates for entropies above $S=200k_B$ with proton to nucleon ratio or electron fraction, $Y_{e}=.30$ and above $S=300k_B$ with $Y_{e}=.40$. However, in these calculations, fission is not particularly important to the abundances at or above the atomic mass range of the rare earth peak until $S\sim275k_B$ for $Y_e=.30$ and $S\sim375k_B$ for $Y_e=.40$.

To account for uncertainties in nuclear physics our nucleosynthesis calculations we use three different nuclear data sets: Finite Range Droplet Model (FRDM) \citep{Moller1995,ADANDT2000,NONSMOKER1998}, Extended Thomas-Fermi with Strutinsky Integral and Quenching (ETFSI-Q) \citep{ETFSIQ1996,ADANDT2000,NONSMOKER1998} and version 17 of the Hartree Fock Bogoliubov model (HFB-17) \citep{HFB17,TALYS}\footnote{http://www.astro.ulb.ac.be/}. We note that the $\beta$-decay rates used for our calculations are only consistent with the FRDM nuclear data set. Since we arrive at our main conclusions after taking averages over and variances due to differences in nuclear models, we expect relatively small quantitative changes in our results from this approximation. However, caution is warranted when attempting to judge the validity of individual nuclear models. The consistency of nuclear data for nuclear astrophysics investigations remains the subject of future research.

\section{Comparing Simulations to Data}\label{sec:analysis}
In this section we compare final abundances to both halo star and solar data. We now comment on each of these data sets.

Spectroscopic observations of galactic halo stars are useful in learning about the nature of nucleosynthesis processes that occur early in the history of the galaxy; see \citet{CS22892} and references within. The consistency of the rare earth elemental abundance data among metal-poor halo stars is quite remarkable. This evidence suggests that these elements were synthesized in the same type of synthesis event \citep{Sneden2008} and by a mechanism which operates over a wide range of metallicity, for review see  \citet{Cowan2006}. Early galactic halo stars also provide an important diagnostic for $r$-process models as the s-process, occurring primarily in lower mass AGB stars, is not believed to have occurred yet.

We compare simulations to an averaged data set of five $r$-process rich halo stars: CS22892-052 \citep{CS22892}, CS31082-001 \citep{CS31082}, HD115444 \citep{HD115444}, HD221170 \citep{HD221170}, and BD+17\textdegree3248 \citep{BD17}. This averaged data set represents a typical $r$-process rich halo star. Before averaging the halo star data we arbitrarily normalized each halo star data to log$\epsilon(Z=63)=0$. We note that this choice of scaling is not critical to the analysis and has been used elsewhere in the literature, e.g. in \citet{Roederer2010}. We denote the averaged halo star elemental abundances as $H(Z)$.

The solar isotopic abundance pattern corresponds to a superposition of many different events during the origin of the solar system \citep{KBW1989,Arlandini1999,Cowan2006,Lodders2009}. A further complication arises when using solar data because in order to extract the solar $r$-process residuals, s-process components must be known accurately \citep{KBW1989,Arlandini1999,Roederer2010}. Despite these intricacies, solar $r$-process elemental yields for the rare earth region are strikingly similar to the elemental abundances of metal-poor, $r$-process rich halo stars. This means that solar data can also be used to constrain favorable freeze-out conditions with the additional advantage of isotopic information. 

In the follow analysis we use the most recent solar data from \citet{Lodders2009}. This data set contains accurately measured abundances among the rare earth elements. To obtain $r$-process residuals, \sdata, we take this solar data and subtract from it the stellar model $s$-process abundances from \citet{Arlandini1999}. The $r$-process residuals, \sdata, are in good agreement with previous data sets \citep{KBW1989,Arlandini1999}.

\subsection{Definitions}\label{sec:defitions}
In this sub section (\ref{sec:defitions}) we introduce a series of definitions that we use in the remainder of section \ref{sec:analysis} to compare simulation to data.

To compare simulations to halo star abundance data, $H(Z)$, we first perform individual calculations using the three nuclear data sets (FRDM, ETFSI, HFB17) with conditions varying between $0\leq n\leq10$, $100k_B\leq S\leq400k_B$, and with $Y_e=.30$ or $Y_e=.40$. We then proceed to take an average of the final abundances from simulations over the different nuclear data holding the conditions constant ($n$,$S$,$Y_e$). We then compare (1) element by element the rare earth peak ($A=159$ to $A=167$ or $Z=64$ through $Z=68$) and (2) the ratio of total abundance in the rare earth peak to the platinum peak ($A=195$ or $Z=78$) of the averaged simulation output to the corresponding values of our compiled halo star.

To accomplish the first comparison we use $G_Z^{REP}$ where,

\begin{equation}\label{eqn:Gz}
G_Z^{REP} \equiv \frac{\sum_{Z=64}^{Z=68} (\alpha Y(Z)-H(Z))^{2}}{\sum_{Z=64}^{Z=68} H(Z)^{2}}
\end{equation}

In equation \ref{eqn:Gz}, $Y(Z) \equiv \sum_A Y(Z,A)$ denotes the final abundance of the stable element with proton number $Z$ and $H(Z)$ is the average abundances of the five halo stars for given proton number. The summation is performed only over the rare earth peak elements as highlighted by the superscript ``REP''. A lower value of $G_Z^{REP}$ implies a better match, with $G_Z^{REP}=0$ representing perfect agreement with $H(Z)$. We scale the final abundance pattern, Y(Z), by $\alpha$ which is the solution to $\frac{d{G_Z^{REP}}}{d\alpha}=0$.

The $A=195$ peak to rare earth peak ratio is an important second constraint as both the lanthanides and heavier elements are believed to be produced in the same synthesis event and so their ratios must match observational data. We compare the ratio of simulations to $H(Z)$ using $\eta_Z$ where,

\begin{equation}\label{eqn:YzRatio}
\eta_Z \equiv \left|{\frac{\alpha Y(Z=78)}{\sum_{Z=64}^{Z=68} \alpha Y(Z)}-\frac{H(Z=78)}{\sum_{Z=64}^{Z=68} H(Z)}}\right|
\end{equation}
Only some of the halo stars have observations of platinum. So for the value of $H(Z=78)$ we average only those halo stars in which this element has been observed.

To compare simulations to the solar $r$-process residuals, \sdata$(A)$, we proceed in a similar fashion using the previous two constraints. As an additional, third, constraint we introduce and measure the amount of late time neutron capture occurring in the rare earth region as these nuclides decay back to stability.

To gauge how close a simulation comes to reproducing the solar rare earth peak we use $G_A^{REP}$,

\begin{equation}\label{eqn:Ga}
G_A^{REP} \equiv \frac{\sum_{A=159}^{A=167} (\alpha Y(A)-N_{\odot,r}(A))^{2}}{\sum_{A=159}^{A=167} N_{\odot,r}(A)^{2}}
\end{equation}
In equation \ref{eqn:Ga}, $Y(A) \equiv \sum_Z Y(Z,A)$ denotes the final abundance along an isobaric chain with atomic mass, $A$ and \sdata$(A)$ is the solar $r$-process residual data also for an isobaric chain with atomic mass, $A$. Again, a lower value of $G_A^{REP}$ implies a better match, with $G_A^{REP}=0$ representing perfect agreement with \sdata$(A)$. The final abundance pattern, Y(A), from simulation is scaled individually by $\alpha$ which is the solution to $\frac{d{G_A^{REP}}}{d\alpha}=0$.

Elemental abundance data from halo stars exhibit an $A=195$ peak to rare earth peak ratio in agreement with the solar elemental abundance distribution. We thus use $\eta_A$,the ratio of the two peak heights for solar data as well.

\begin{equation}\label{eqn:YaRatio}
\eta_A \equiv \left|{\frac{\sum_{A=190}^{A=200} \alpha Y(A)}{\sum_{A=159}^{A=167} \alpha Y(A)}-\frac{\sum_{A=190}^{A=200} N_{\odot,r}(A)}{\sum_{A=159}^{A=167} N_{\odot,r}(A)}}\right|
\end{equation}

Our third constraint comes from the observation that too much neutron capture during freeze-out has negative consequences for the final $r$-process abundances \citep{Arcones2011a,Mumpower2011}. This is particularly important for the rare earth region because the neutron capture rates in this region are on average faster than in other regions during the last stage of the $r$-process \citep{Arcones2011a,Mumpower2011}. Late-time neutron capture effect shifts material from below the peak ($A=150$ to $A=158$) to above the peak ($A=168$ to $A=180$) causing an over production of the heavier rare earth elements, which does not resemble either the solar $r$-process residuals, or halo star observations.

\begin{figure}
      \includegraphics[width=160mm,height=58mm]{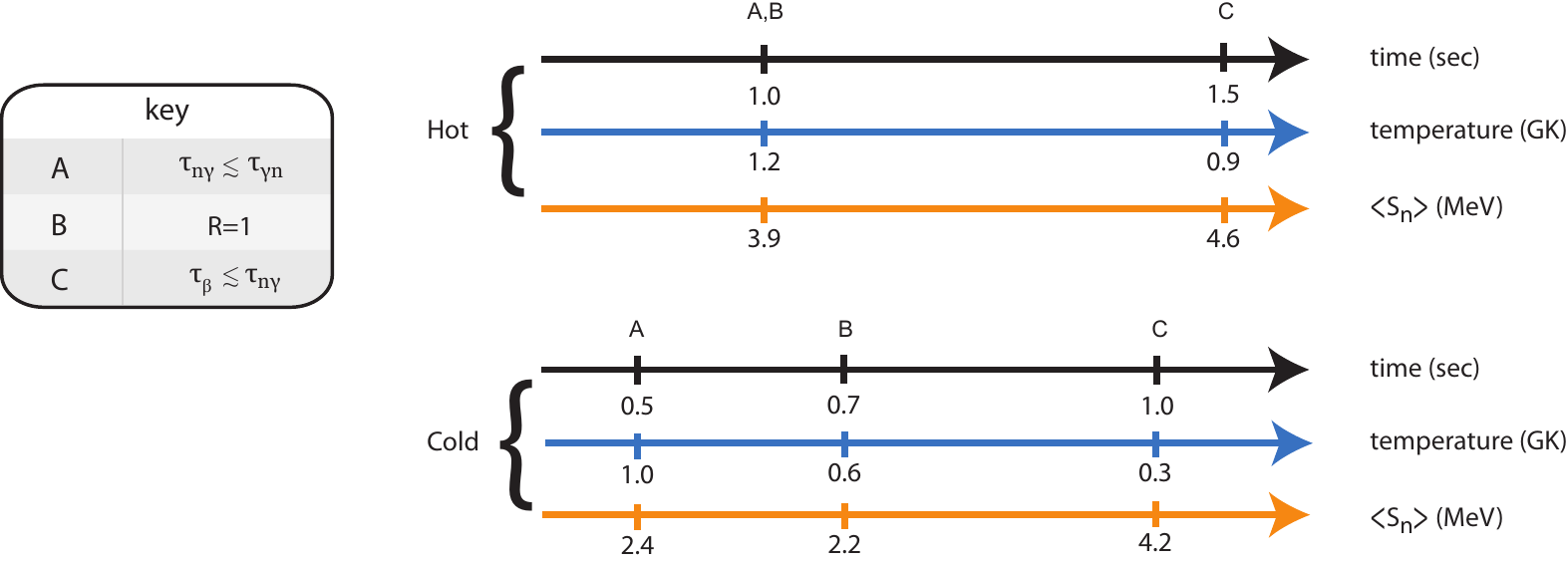}
      \caption{\label{fig:timeline} Contrasts the timeline of important events during the formation and evolution of the rare earth region between hot and cold evolutions. Point A represents the beginning of \nggn \ freeze-out, point B neutron exhaustion (neutron-to-seed ratio of unity, $R=1$), and point C when the timescale for $\beta$-decays take over neutron capture.}
\end{figure}

\begin{figure}
   \begin{center}
      \includegraphics[width=160mm,height=60mm]{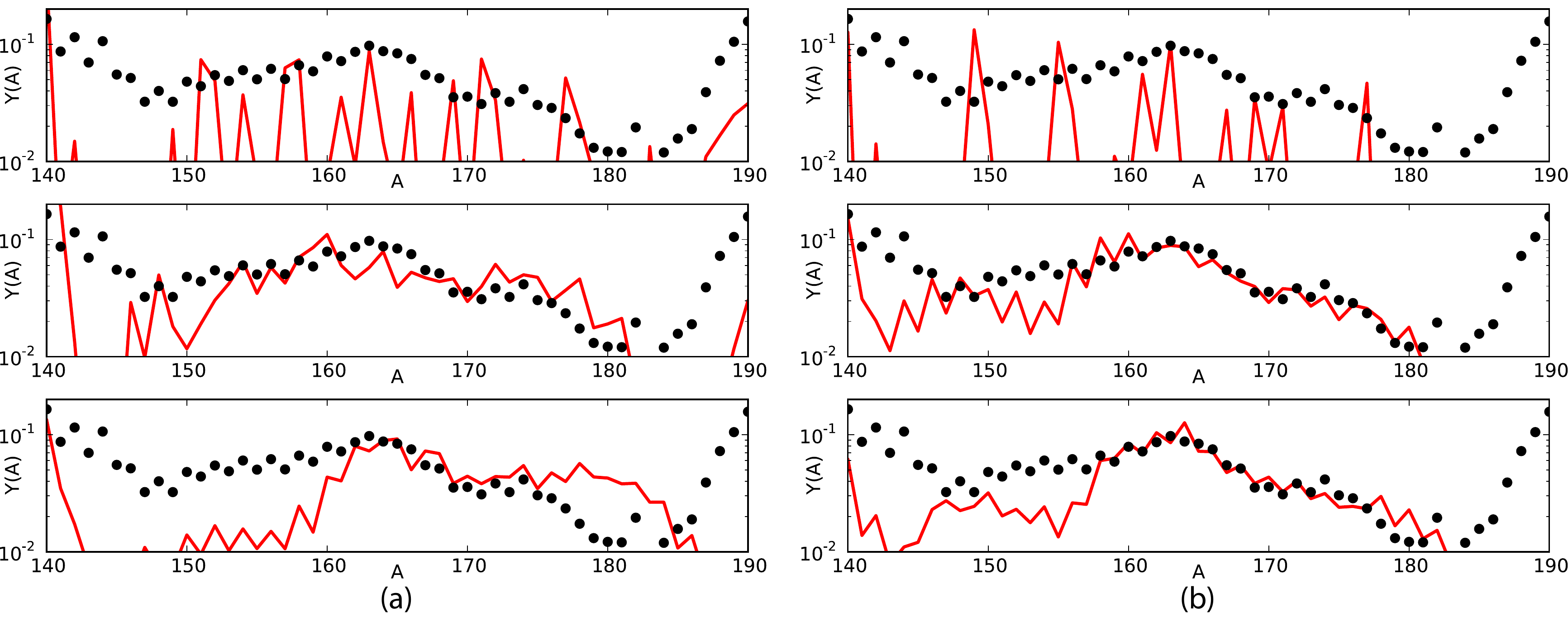}
      \caption{\label{fig:tmnc}  Highlights the late time neutron capture effect by showing the time evolution of rare earth region abundances versus the solar $r$-process curve (black) at three snapshots during the simulation. Panel (a) shows an $r$-process where late time neutron capture occurs using conditions ($n=6.0$,$S=350k_B$,$Y_e=.30$). Panel (b) shows little to no late time neutron capture using conditions ($n=2.0$,$S=350k_B$,$Y_e=.30$). Calculations were performed with the FRDM nuclear data. Abundances in the top row are shown at $R=1$. Abundances in the middle row are shown in between $R=1$ and the point at which $\beta$-decays take over. The bottom row shows final abundances. Refer back to Figure \ref{fig:timeline}.}
   \end{center}
\end{figure}

We contrast the abundances of two simulations to exhibit the late-time neutron capture effect in Figure \ref{fig:tmnc}. In both panels of the figure we show a snapshot of abundances at different points during the calculation. The top panel shows abundances at $R=1$ (see point A, Figure \ref{fig:timeline}), the middle panel shows abundances during freeze-out (in between point B and C, Figure \ref{fig:timeline}) and the bottom panel shows the final isotopic abundances as compared to the solar abundances. The peak forms successfully in both cases, panel (a) with conditions ($n=6.0$,$S=350k_B$,$Y_e=.30$) and panel (b) with conditions ($n=2.0$,$S=350k_B$,$Y_e=.30$). However, comparing the middle and bottom snapshots in panel (a) we can clearly see the late time neutron capture effect during freeze-out. This effect is not present when comparing the same snapshots in panel (b).

The late-time neutron capture effect is especially useful because, as we have just shown, it too is highly dependent on the conditions at late-times. We thus can use this effect as a constraint on favorable $r$-process conditions. To estimate the amount of late time neutron capture which shifts material across the rare earth region we compute the difference between the value of $G_A$ during freeze-out (between points B and C on the timeline) and the value of $G_A$ using the final abundances. The summation now occurs over all rare earth elements ($A=150$ to $A=180$) signified by the superscript ``REE''.

\begin{equation}\label{eqn:DeltaGa}
\Delta G_A^{REE} \equiv G_A^{REE}(\text{final})-G_A^{REE}(\text{freeze-out})
\end{equation}

\subsection{Determining Freeze-out Conditions}
In Figure \ref{fig:yz-contour} we highlight astrophysical conditions where simulations best match $H(Z)$. The power law ($n$) from the density parameterization is plotted on the y-axis along with entropy per baryon ($S$) on the x-axis. Panel (a) simulations have $Y_{e}=.30$ and panel (b) simulations have $Y_{e}=.40$. Shown in each panel is a solid vertical line which denotes the point at which a sufficient neutron-to-seed ratio ($R\sim80$ at $T_9\sim3$) has been reached to produce a full $r$-process with elements out to $A\sim200$ \citep{HWQ1997,MeyerBrown1997,TBM2001}. We also denote the point at which fission cycling influences the final abundances of the $A\gtrsim150$ region by a dotted vertical line. 

In Figure \ref{fig:ya-contour} we highlight astrophysical conditions where simulations best match \sdata$(A)$. The axes and markings are the same as in Figure \ref{fig:yz-contour}. The left column shows regions of parameter space calculated with $Y_{e}=.30$ and the right column with $Y_{e}=.40$.

We now discuss below each of the abundance features individually and their impact on constraining freeze-out conditions when comparing to halo star and solar data.

\subsubsection{Using Rare Earth Peak Formation}
Shown in both panels of Figure \ref{fig:yz-contour} is a dark shaded region (online red) where the scaled final abundances of rare earth peak elements from simulations best agree with the abundances of rare earth peak elements of $H(Z)$. Since no single simulation reproduces exactly the $H(Z)$ abundances, we must compute an appropriate cut-off for $G_Z^{REP}$. To this end, we measure the variance among the nuclear models by computing the standard deviation of the individual abundances at each value of $Y(Z)$ given constant conditions ($n$,$S$,$Y_e$). We additionally estimate the uncertainty in astrophysical environment by computing a second average over the simulations. Propagation of error from averaging the nuclear models and averaging over astrophysical conditions produces a value of $G_Z^{REP}=0.03$. We thus take this value as a cut-off for shading the contour.

Shown in both columns of panel (a) of Figure \ref{fig:ya-contour} is a dark shaded region (online red) where the scaled final abundances of rare earth peak elements from simulations best agree with the abundances of rare earth peak elements of \sdata$(A)$. We repeat the same procedure as mentioned above to compute the cut-off value of $G_A^{REP}=0.15$ for shading this contour.

Fission during freeze-out operates under higher entropy components ($S>200k_B$ for $Y_e=.30$ and $S>300k_B$ for $Y_e=.40$) and influences abundances at or above the rare earth region at $S\sim275k_B$ for $Y_e=.30$ and $S\sim375k_B$ for $Y_e=.40$. We denote the point at which fission is influential in both Figures \ref{fig:yz-contour} and \ref{fig:ya-contour} by a dotted line in each case. Without fission, the dark shaded contour would end at or before the dotted line. As can be seen from the figures, the inclusion of fission extends the favorable calculated freeze-out conditions for rare earth peak formation.

\subsubsection{Using The A=195 Peak to Rare Earth Peak Ratio}
The light shaded region (online yellow) in both panels of Figure \ref{fig:yz-contour} represents the parameter space in which simulations produce the same ratio of the $A=195$ peak to rare earth peak as compared to the averaged halo star, $H(Z)$. We take as an upper bound to this contour, $\eta_Z=2$, which represents an over production of the $A=195$ peak relative to the rare earth peak by a factor of two, the value of $\eta_Z\approx2$ comes from the ratio $\frac{H(Z=78)}{\sum_{Z=64}^{Z=68} H(Z)}$. We note that the under production of the $A=195$ peak is taken care of by the sufficient neutron-to-seed ratio constraint (solid vertical line). All together, this results in a constraint similar to the method used by \citet{MeyerBrown1997}. Qualitatively similar behavior is shown in the left and right columns of panel (b) of Figure \ref{fig:ya-contour}, where calculated final abundances are compared to solar data. We take as an upper bound to this contour, $\eta_A=4$ which again represents an over production of the $A=195$ peak relative to the rare earth peak by approximately a factor of two.

Interestingly, the left column of both figures, those with $Y_e=.30$, show a second favorable region for this ratio at higher entropy. This is due to the oscillation of the $A=195$ peak as a function of $R$ due to fission cycling as discussed in \citet{Beun2006,Beun2008}. In the gap, between the two favorable regions the $A=195$ peak is over produced by more than a factor of two compared to the rare earth peak. This effect also occurs in the  $Y_e=.40$ case, however our simulations stop at $S=400k_B$ and so the effect is not shown.

\subsubsection{Using The Late-Time Neutron Capture Effect}
Returning to panel (c) of Figure \ref{fig:ya-contour}, the shaded region (online green) in both columns represents the region of parameter space in which simulations do not exhibit a late-time neutron capture effect. We take as an upper bound to this contour, $\Delta G_A^{REE}=1$, which represents significant fluctuations occurring across the rare earth region during freeze-out.

Inspection of panel (c) reveals that the late-time neutron capture effect is sensitive to freeze-out conditions, this is particularly evident at high entropy in the left column with $Y_e=.30$. During a cold freeze-out (large values of $n$) the path moves far from stability where $\beta$-delayed neutron probabilities are large. As the nuclides decay back to stability, $\beta$-delayed neutron reactions provide an ample supply of neutrons for capture. Additionally, fission actively introduces new seed nuclei into the region. Hence, the late-time neutron capture effect returns in this section of the figure causing the shaded contour to disappear. At high entropy, in a hot freeze-out (lower values of $n$), the late-time neutron capture effect does not occur (presence of the shaded region) due to the presence of photo-dissociation and prolonged \nggn \ equilibrium.

\subsection{Further Analysis}
\begin{figure}
   \begin{center}
      \includegraphics[width=160mm,height=70mm]{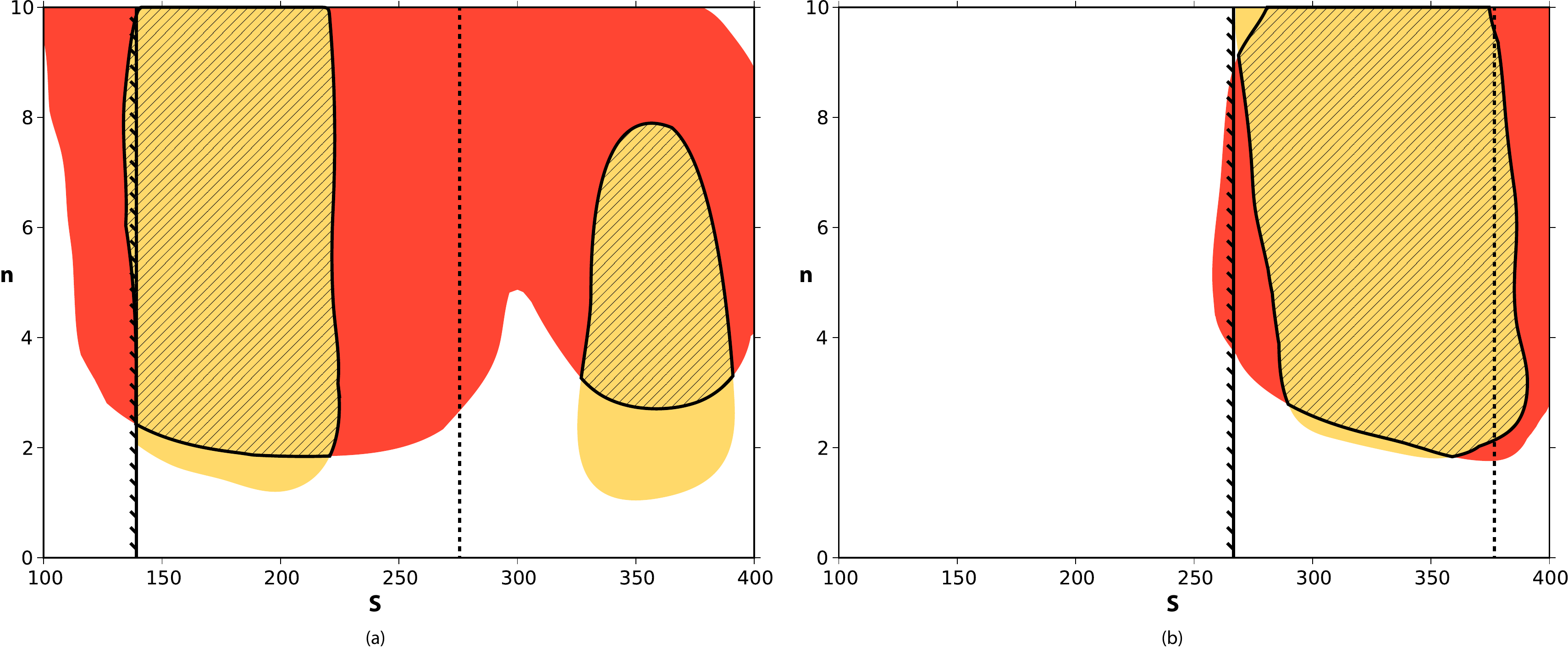}
      \caption{\label{fig:yz-contour}  Shows how the rare earth region can be used to constrain astrophysical conditions when comparing simulations to typical halo star data. Calculations were performed with $Y_e=.30$, panel (a) and $Y_e=.40$, panel (b). The dark contour (online red) highlights conditions which produce a rare earth peak which best matches the average of five halo stars, $H(Z)$. The light contour (online yellow) highlights conditions which produce an $A=195$ peak to rare earth peak which best matches $H(Z)$. The intersection of these constraints is showing by the hatching. A full $r$-process out to $A\sim200$ occurs above entropies denoted by the solid vertical line. Above the dotted line fission influences the abundances of $A\gtrsim150$ nuclides.}
   \end{center}
\end{figure}

\begin{figure}
   \begin{center}
      \includegraphics[width=120mm,height=171.43mm]{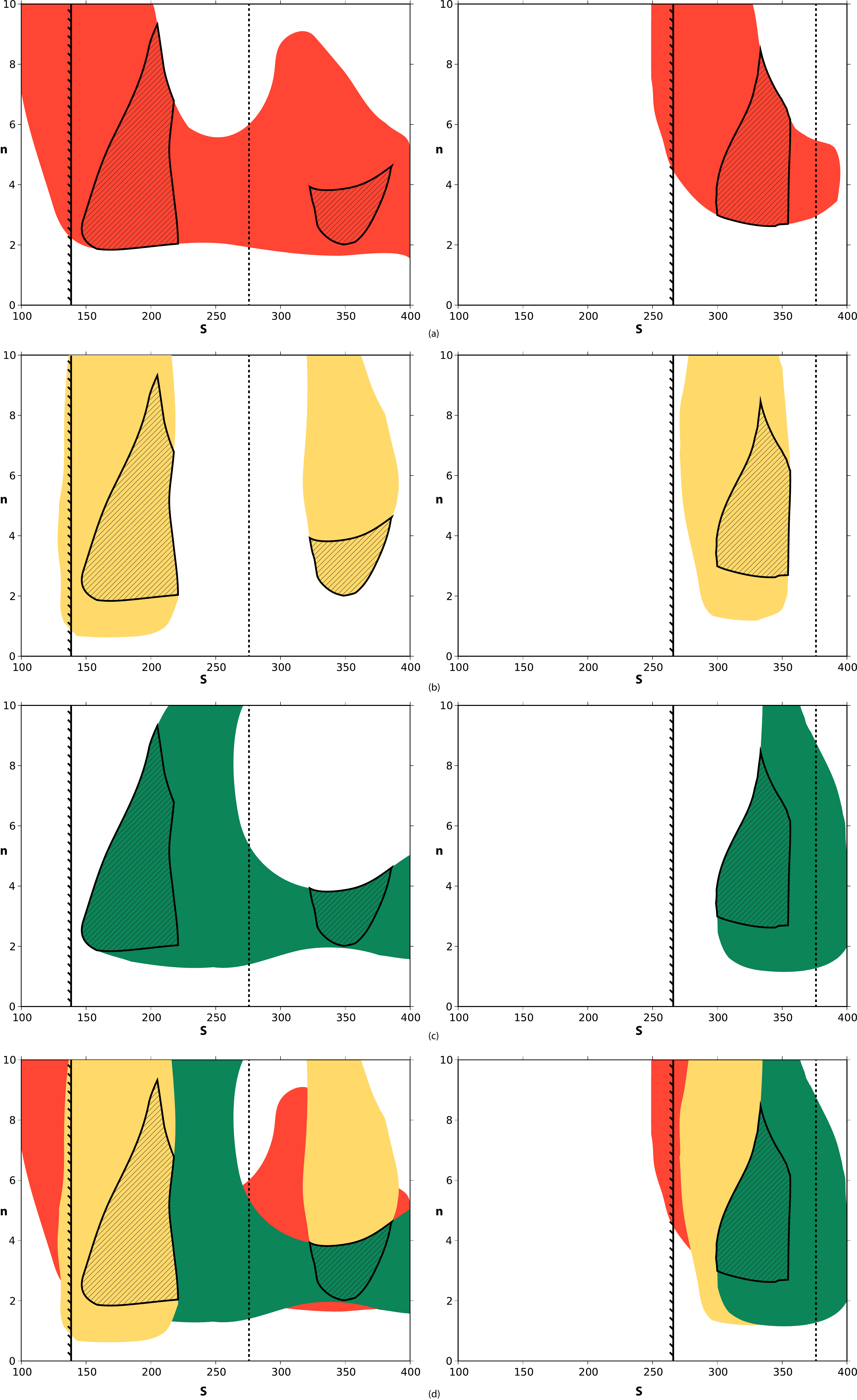}
      \caption{\label{fig:ya-contour}  Shows how the rare earth peak can be used to constrain astrophysical conditions when comparing simulations to the solar $r$-process data. The axes are the same as Fig. \ref{fig:yz-contour}. The left column displays computations performed with $Y_{e}=.30$ and right column with $Y_{e}=.40$. Panel (a) highlights regions (online red) where the rare earth peak elements $A=159$ to $A=167$ from simulations best agree with the rare earth peak elements of the solar $r$-process data. Panel (b) shows regions (online yellow) where simulations produce the same ratio of the $A=195$ peak to rare earth peak as compared to the solar data. Panel (c) highlights regions (online green) where the late time neutron capture effect is minimal. The intersection (shaded region) of these three constraints are highlighted in each panel and repeated in Panel (d) with all three constraints. A full $r$-process out to $A\sim200$ occurs above entropies denoted by the solid vertical line. Above the dotted line fission influences the abundances of $A\gtrsim150$ nuclides.}
   \end{center}
\end{figure}

In order to emphasize the differences between the classical neutron-to-seed ratio constraint (the region of parameter space above the solid vertical line) and constraints using the rare earth elements we highlight in each panel of Figures \ref{fig:yz-contour} and \ref{fig:ya-contour} a hatched region which represents our final constraint.

In Figure \ref{fig:yz-contour}, where simulation is compared with halo star data, the final constraint region is generally controlled by the $A=195$ peak to rare-earth-peak ratio as the rare earth peak formation alone is a less stringent constraint. In Figure \ref{fig:ya-contour}, where simulation is compared with solar data, the final constraint region is generally controlled by the same peak height ratio and additionally the late-time neutron capture effect.

When comparing to both meteoritic and observational data given an initial $Y_e=.30$, we find the entropy range favorable for rare earth peak formation is $S\sim150k_B$ to $S\sim200k_B$ without fission. While the rare earth peak is formed by two different mechanisms in this entropy range, it does not favor one mechanism over the other. With the inclusion of fission a second favorable region appears around $S\sim325k_B$ to $S\sim375k_B$ in both figures. In this region the rare earth peak formation and peak height constraints rule out the warmest and coldest scenarios when comparing to halo star data. When comparing to halo star data at high entropy, hot conditions are disfavored due to the constraint on the rare earth peak elements and colder conditions are disfavored because the $A=195$ peak becomes over produced. When comparing to solar data, the late-time neutron capture effect rules out colder freeze-out scenarios (see discussion in previous section).

We now explicitly investigate final abundances in the $Y_e=.30$ case to the halo star and solar data. In order to observe the abundance pattern which is typical of the region suggested by the new constraint procedure we average the final abundances in the constrained region and compute the standard deviation to measure the variance of these final abundances.

\begin{figure}
   \begin{center}
      \includegraphics[width=160mm,height=120mm]{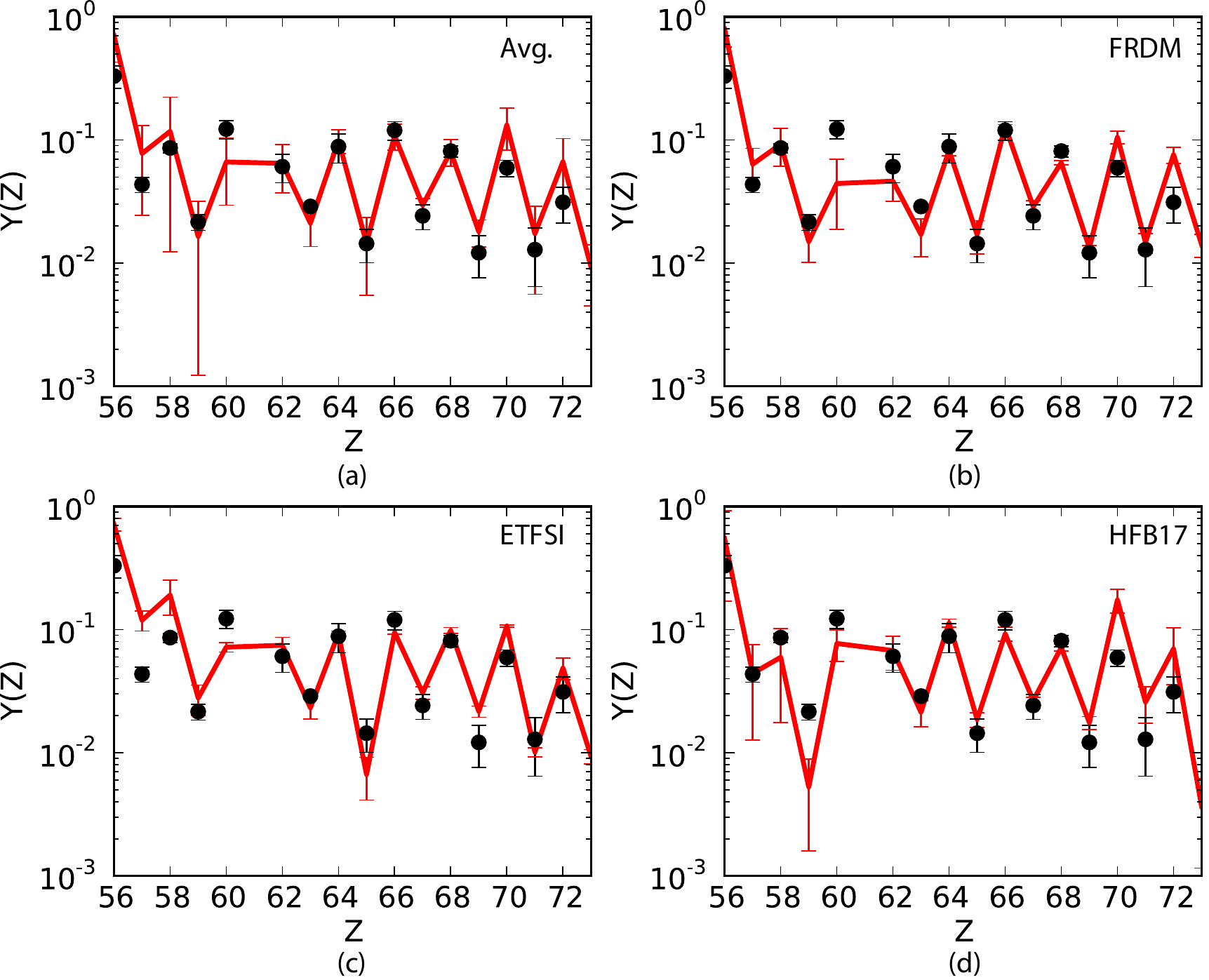}
      \caption{\label{fig:yz-final}  Shows calculated final averaged elemental abundance patterns (solid red) along with uncertainties compared to the average of five halo stars, $H(Z)$ (dots). All calculations were performed with $Y_e=.30$. Panel (a) shows an average over 3 nuclear data sets and astrophysical conditions from the constraint region of Figure \ref{fig:yz-contour}. Panels (b)-(d) show the average over astrophysical conditions in the constraint region for each individual nuclear model.}
   \end{center}
\end{figure}

\begin{figure}
   \begin{center}
      \includegraphics[width=160mm,height=120mm]{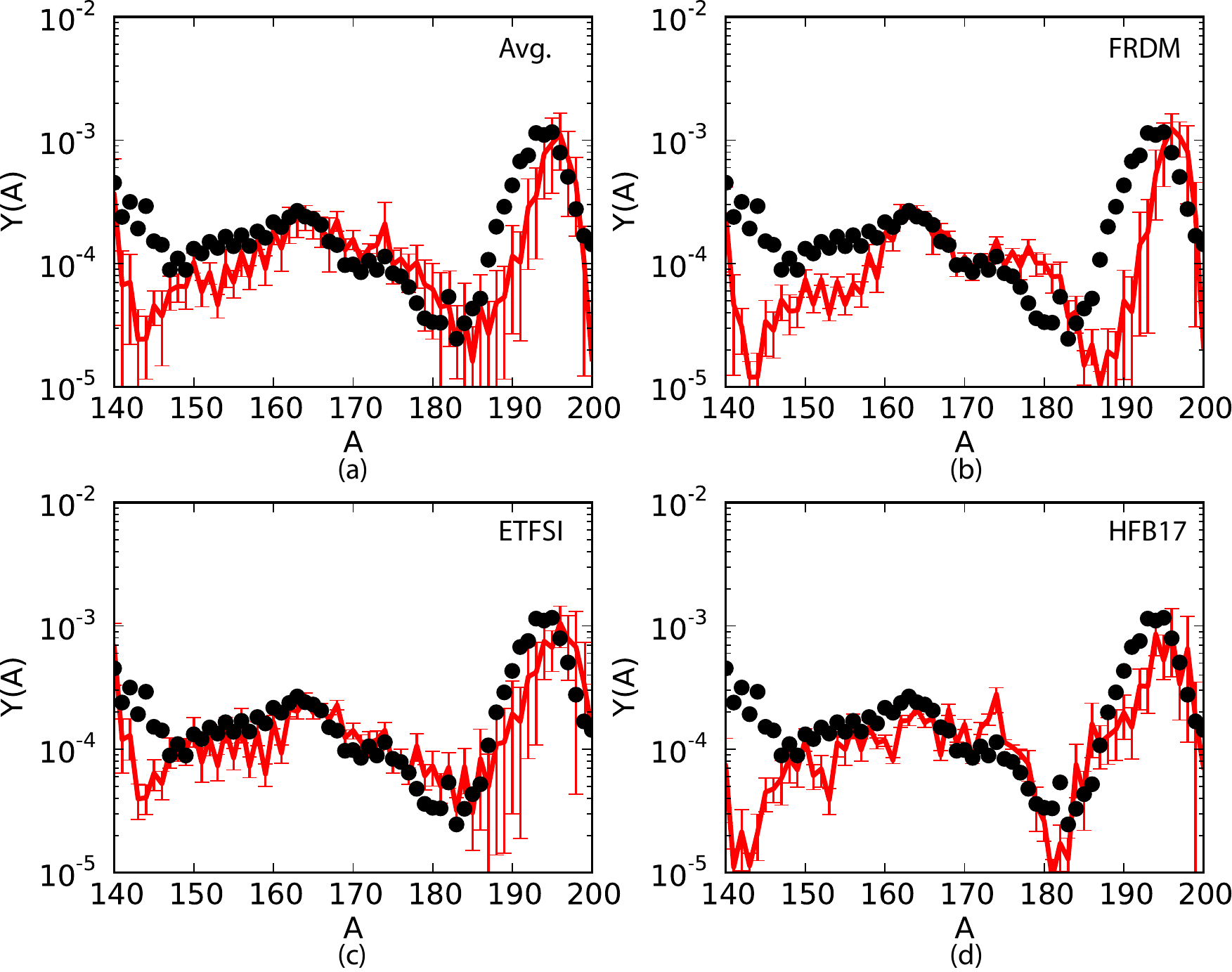}
      \caption{\label{fig:ya-final}  Shows calculated final averaged isotopic abundance patterns (solid red) along with uncertainties compared to the solar $r$-process residuals (black dots). All calculations were performed with $Y_e=.30$. Panel (a) shows the average over 3 nuclear data sets and astrophysical conditions from the constraint region of Figure \ref{fig:ya-contour}. Panels (b)-(d) show the average over astrophysical conditions in the constraint region for each individual nuclear model.}
   \end{center}
\end{figure}

The results of these calculations are shown in Figure \ref{fig:yz-final} and \ref{fig:ya-final}. Panel (a) includes both averages over nuclear physics data sets and astrophysical conditions and the error bars represent variance from nuclear models and astrophysical conditions. The abundances in panel (a) correspond with simulations in the hatched region of the left column of Figure \ref{fig:yz-contour} and \ref{fig:ya-contour} respectively. Panels (b)-(d) show the final abundances of individual nuclear models after averaging over their respective constraint regions and the error bars represent the variance from astrophysical conditions alone.

As can be seen from Figure \ref{fig:yz-final}, these simulations do very well in reproducing the rare earth peak. When comparing to $H(Z)$, the largest variation in the rare earth peak region comes from Terbium, $Z=65$. Among all of the lanthanides, Praseodymium, $Z=59$, shows the largest variance while Neodymium, $Z=60$, is consistently under produced. Figure \ref{fig:ya-final} also shows similar results for the rare earth peak elements. When comparing to the solar data, the largest variances in the final abundances can be found in the $A\sim180$ to $A\sim195$ region. This effect has been suggested to arise from a lack of long range correlations in the nuclear models \citep{Arcones2011b}.

It is important to note in both figures that the variance seen in the final abundances of the rare earth elements is small. While subtleties persist in the individual nuclear models, global trends consistent with halo star and solar abundances distribution are present. As expected, we isolate conditions where the rare earth peak and the solar $A=195$ to rare earth peak ratio mimic the solar data and further, prevent too much build up of the heavier rare earths. The success of this method is clearly dependent on how well we understand the input nuclear physics.

\section{Summary and Conclusions}
Using the delicate, out of \nggn \ equilibrium formation process of the rare earth peak we have studied and isolated freeze-out conditions in high entropy ($S\gtrsim100k_B$) $r$-process calculations conducive for production of these elements. We compare the final abundances from simulations to both the elemental abundances of halo stars and the isotopic abundances of the solar $r$-process residuals. Both of these data sets suggest similar astrophysical conditions during freeze-out. For example, without fission, we find that favorable freeze-out conditions fall within a narrow entropy window ($S\sim150k_B$ to $S\sim200k_B$) given an initial electron fraction, $Y_e=.30$ and timescale of $\tau\sim27 \text{ms}$. Given an initial electron fraction, $Y_e=.40$ and same initial timescale we find an entropy window of $S\sim300k_B$ to $S\sim350k_B$.

The sensitivity of final abundances to uncertainties in nuclear physics is considered by including calculations with the FRDM, ETFSI and HFB17 nuclear models. We consider the variance among final abundances due to these three models and include it when determining the appropriate constraint for the formation of the rare earth peak (values of $G_Z^{REP}$ and $G_A^{REP}$).

We use three abundance features to constrain astrophysical conditions: (1) We require the formation and correct shape of the rare earth peak ($A=157$ to $A=167$ or $Z=64$ to $Z=68$). The importance of this process and its sensitivity to freeze-out conditions was discussed in \citet{Mumpower2011}. (2) We utilize the constraint on the ratio of the abundance in the rare earth peak to that of the $A=195$ peak, as these elements are believed to be synthesized in the same environment. (3) We utilize a new constraint on freeze-out behavior: the late-time neutron capture constraint.

The late-time neutron capture constraint involves studying the shifting of material between isotopes during freeze-out in the rare earth region. We find the over production of the heaviest rare earth elements is highly dependent on the conditions during freeze-out, making it a valuable consideration for $r$-process models. This effect can be measured by comparing abundances to solar $r$-process residuals as this data provides isotopic information.

The impact of fission during freeze-out is to produce multiple islands of favorable $r$-process conditions, see e.g. the left column of Figures \ref{fig:yz-contour} or \ref{fig:ya-contour}. The late-time neutron capture constraint becomes tighter when fission occurs. This is due to two factors: (1) A cold $r$-process freeze-out extends far from stability where $\beta$-delayed neutron probabilities are large, providing additional free neutrons to capture during freeze-out. (2) The inclusion of fission during freeze-out cycles additional material through the rare earth region which can capture the free neutrons. This suggests that if fission is important for the formation of the $r$-process elements, the astrophysical freeze-out conditions are more likely to be `hot.'

We have shown that the formation of the rare earth peak is a useful tool in determining favorable $r$-process freeze-out conditions. This diagnostic works well despite current uncertainties in nuclear physics. It complements the commonly used neutron-to-seed ratio and should be used in addition when analyzing astrophysical models. The size of the range of acceptable astrophysical conditions is dominated by uncertainty in the nuclear physics inputs. Reduction of this uncertainty through an improved understanding of the properties of nuclei far from stability and self-consistent nuclear data will increase the efficacy of this tool.

\section{Acknowledgements}
We thank Raph Hix for providing the charged particle network and up-to-date reaction libraries. 
We thank North Carolina State University for providing the high performance computational resources necessary for this project. 
This work was supported in part by U.S. DOE Grant No. DE-FG02-02ER41216, DE-SC0004786, and DE-FG02-05ER41398.

\bibliographystyle{unsrtnat}

\clearpage

\end{document}